\documentclass{moriond}

\def\Journal#1#2#3#4{{#1} {\bf #2}, #3 (#4)}
\def\etal{{\em et al.}}

\def\NPB{{\em Nucl. Phys.} B}
\def\PLB{{\em Phys. Lett.}  B}

\def\PRD{{\em Phys. Rev.} D}

\def\ra{\rightarrow}

\def\beq{\begin{equation}}
\def\eeq{\end{equation}}
\def\bea{\begin{eqnarray}}
\def\eea{\end{eqnarray}}

\newcommand{\Eq}[1]{Eq.~(\ref{eq#1})}
\def\pmdiff#1#2{\raise.5ex\hbox{$\sss +#1}$%
    \kern-2.8em\lower1ex\hbox{${\sss-#2}$}} 
\def\decayarrow{\kern0.2em\hbox{$\raise1.08ex\hbox{\big|}\kern-0.5em
                \longrightarrow$}}   

\newcommand{\Photo}{}

\begin{document}
\vspace*{2cm}
\title{ADDRESSING THE MAJORANA VS. DIRAC QUESTION USING NEUTRINO DECAYS
\footnote{To appear in the Proceedings of the 53$^\mathrm{rd}$ Rencontres de Moriond Electroweak session, held in March, 2018.
\newline Fermilab report FERMILAB-CONF-18-186-T.
\newline Based on A. B. Balantekin and B. Kayser, submitted to \textit{Ann.\ Rev.\ Nucl.\ Part.\ Sci.,} arXiv:1805.00922 [hep-ph], and on A. B. Balantekin, A. de Gouv\^{e}a, and B. Kayser, in preparation.}}

\author{ BORIS KAYSER }
\address{Theoretical Physics Department, Fermilab, P.O. Box 500, Batavia, IL 60510  USA}

\maketitle\abstracts{
We explain why it is so hard to determine whether neutrinos are Majorana or Dirac particles as long as the only neutrinos we study are ultra-relativistic. We then show how non-relativistic neutrinos could help, and focus on the angular distributions in the decays of an as-yet-to-be-discovered heavy neutrino $N$. We find that these angular distributions could very well tell us whether neutrinos are Majorana or Dirac particles.}


One of the most basic questions about the neutrinos is whether every neutrino mass eigenstate is a Majorana particle (that is, identical to its antiparticle), or a Dirac particle (that is, distinct from its antiparticle). Determining whether neutrinos are Majorana or Dirac particles experimentally is very challenging. To understand why, let us note first that all the neutrinos we have been able to study directly so far have been ultra-relativistic. As an example, let us consider the neutrinos from pion decay. The decay $\pi^+ \ra \mu^+ + \nu_\mu$ produces a neutrino $\nu_\mu$ that is not only ultra-relativistic but also of essentially 100\% left-handed helicity. Correspondingly, the decay $\pi^- \ra \mu^- + \overline{\nu_\mu}$ produces an ultra-relativistic neutral lepton that is an antineutrino if indeed antineutrinos are distinct from neutrinos, and that is of essentially 100\% right-handed helicity.
Now, suppose the neutral lepton from a $\pi \ra \mu \nu$ decay undergoes a charged-current weak interaction with some target, producing an outgoing muon in the process. The Standard Model (SM) Lagrangian density describing this interaction is
\beq
{\cal L}_{CC} \propto \bar{\mu} \gamma^\lambda \frac{(1-\gamma_5)}{2} \nu_\mu J_\lambda + 
	\overline{\nu_\mu} \gamma^\lambda \frac{(1-\gamma_5)}{2} \mu J_\lambda^\dag \;\; ,
\label{eq1}
\eeq
where $J_\lambda$ is a current that pertains to the target. As we know, the field $\mu$ in this Lagrangian creates only a $\mu^+$, not a $\mu^-$, while the field $\bar{\mu}$ creates only a $\mu^-$, not a $\mu^+$. Similarly, {\em in the Dirac case}, the field $\nu_\mu$ absorbs only a neutrino, while $\overline{\nu_\mu}$ absorbs only an antineutrino. Thus, in the Dirac case, only the first term on the right-hand side of \Eq{1} can absorb the neutral lepton from the decay $\pi^+ \ra \mu^+ + \nu_\mu$, so only a $\mu^-$, not a $\mu^+$, can be produced in the interaction. Similarly, only the second term can absorb the neutral lepton from $\pi^- \ra \mu^- + \overline{\nu_\mu}$, so only a $\mu^+$, not a $\mu^-$, can be produced. The lepton number $L$ that distinguishes antileptons from leptons is conserved.

In the Majorana case, both the fields $\nu_\mu$ and $\overline{\nu_\mu}$ can absorb a neutrino. However, owing to the left-handed chiral projection operator $(1-\gamma_5)/2$, when the neutrino is ultra-relativistic in the rest frame of the target, only the first (second) term on the right-hand side of \Eq{1} can absorb it if it is of $\sim 100\%$ left-handed (right-handed) helicity. Thus, only a $\mu^- \; (\mu^+)$ will be produced if the neutrino is from $\pi^+ \; (\pi^-)$ decay. That is, the result of the interaction with the target will be identical to what it is in the Dirac case.

As this example illustrates, in almost all circumstances, when neutrinos are ultra-relativistic, helicity is a substitute for the lepton number $L$. Whether there is a conserved lepton number (Dirac case) or not (Majorana case) makes no practical difference. Majorana and Dirac neutrinos behave indistinguishably. 

In contrast, {\em non-relativistic} Majorana and Dirac neutrinos can behave quite differently. To illustrate, let us consider an electron neutrino that is non-relativistic in the rest frame of some target. Suppose this neutrino initiates on the target an exothermic reaction in which a charged lepton is produced. The SM Lagrangian for this reaction, similar to that in \Eq{1}, is 
\beq
{\cal L}_{CC} \propto \bar{e} \gamma^\lambda \frac{(1-\gamma_5)}{2} \nu_e J_\lambda + 
	\overline{\nu_e} \gamma^\lambda \frac{(1-\gamma_5)}{2} e J_\lambda^\dag \;\; .
\label{eq2}
\eeq
If the incoming neutrino is a Dirac particle tagged as a neutrino rather than an antineutrino by the process that created it, it can be absorbed only by the first term on the right-hand side of \Eq{2}, and consequently it can produce only an electron, not a positron. However, if it is a Majorana neutrino, then regardless of how it was created, and regardless of its helicity, it can be absorbed by either of the terms on the right-hand side of \Eq{2}. The left-handed chiral projection operator $(1-\gamma_5)/2$ does not significantly suppress the absorption of non-relativistic neutrinos of either helicity by either of these two terms. Thus, a Majorana  neutrino can produce either an electron or a positron.

The observation that non-relativistic Majorana and Dirac neutrinos can behave quite differently leads us to wonder if nature contains a so-far-undiscovered heavy neutrino $N$ whose decays could be studied. After all, in its rest frame---the natural frame in which to consider its decays---a neutrino is completely non-relativistic. The observation at a hadron collider of a lepton-number-nonconserving sequence such as
\begin{eqnarray}
\mathrm{quark + antiquark} \ra \;\;\; W^+ &\ra & N + \mu^+  \;\;\;\;\;\;\; ,  \nonumber   \\
	& &  \decayarrow \; e^+ + \pi^-   
\label{eq3}
\end{eqnarray}
which is forbidden if $N$ is a Dirac particle, would signal that the neutrinos, including $N$, are Majorana particles. However, if the $N$ is created at a neutrino oscillation experiment, lepton-number violation such as in the sequence (3) may be impossible to detect because the detector may not be able to tell whether a charged particle is positive or negative.    
Thus, it is quite interesting that the Majorana or Dirac character of neutrinos could also be revealed by the angular distribution of the particle $X$ in a decay of the form $N \ra \nu + X$. Here, $\nu$ is one of the established light neutrino mass eigenstates, $\nu_1, \; \nu_2$, or $\nu_3$, and $X$ is a self-conjugate boson: $\bar{X} = X$. Depending on the mass of $N, \; X$ could be, for example, a $\gamma, \; \pi^0, \; \rho^0, \; Z^0$, or the Higgs boson $H^0$, and we shall consider these five cases \cite{ref0}.

For each of the decay modes under consideration, the decay rate $\Gamma(N \ra \nu + X)$ will be twice as big if $N$ and $\nu$ are Majorana particles as it will be if they are Dirac particles \cite{ref1}. However, this difference may not be too useful, because the decay rate also depends on unknown mixing angles. Therefore, we turn to the decay angular distribution in the $N$ rest frame. We assume that the mechanism that produces the $N$ leaves it fully polarized, with its spin vector $\vec{s}$ pointing in a space-fixed direction we shall call the $+z$ direction. 
We denote the $X$ and $\nu$ helicities by $\lambda_X$ and $\lambda_\nu$, respectively, and define $\lambda \equiv \lambda_X - \lambda_\nu$. The quantity $\lambda$ is the projection $\vec{J}_{\mathrm{final}} \cdot \hat{p}$ of the total final-state angular momentum $\vec{J}_{\mathrm{final}}$ along the direction $\hat{p}$ of the outgoing particle $X$.

From rotational invariance alone, it follows that the differential decay rate for $N \ra \nu + X$ is given as a function of the angle $\theta$ between $\hat{p}$ and $\vec{s}$ by
\begin{eqnarray}
\frac{d \Gamma(N \ra \nu + X)}{d(\cos \theta)} & = & \frac{\Gamma_{\lambda = +1/2}}{2} (1 + \cos \theta) + \frac{\Gamma_{\lambda = -1/2}}{2} (1 - \cos \theta)  \nonumber \\
	& = & \frac{\Gamma_0}{2} ( 1 + \alpha \cos \theta) \; ; \; -1 \leq \alpha \leq +1 \;\; . 
\label{eq4}
\end{eqnarray}
Here, $\Gamma_{\lambda = +1/2}$ and $\Gamma_{\lambda = -1/2}$ are the rates for decay into all the final states with $\lambda = + 1/2$, and all those with $\lambda = - 1/2$, respectively. 
$\Gamma_0 = \Gamma_{\lambda = +1/2} + \Gamma_{\lambda = -1/2}$ is the total decay rate for $N \ra \nu + X$, and $\alpha = (\Gamma_{\lambda = +1/2} - \Gamma_{\lambda = -1/2}) /( \Gamma_{\lambda = +1/2} + \Gamma_{\lambda = -1/2})$ is the asymmetry parameter for this decay.

To an excellent approximation, the heavy neutrino decay $N \ra \nu + X$ is described by an amplitude $\langle X(\hat{p},\lambda_X)\, \nu(-\hat{p},\lambda_\nu) | H | N(\vec{s}) \rangle$ that is first order in some Hermitean Hamiltonian $H$. If $N$ and $\nu$ are Majorana particles, so that every participant in the decay $N \ra \nu+X$ is a self-conjugate particle, CPT plus rotational invariance implies that
\beq
|\langle X(\hat{p},\lambda_X)\, \nu(-\hat{p},\lambda_\nu) | H | N(\vec{s}) \rangle |^2  =
|\langle X(-\hat{p},-\lambda_X)\, \nu(\hat{p},-\lambda_\nu) | H | N(\vec{s}) \rangle |^2 \;\; .
\label{eq5}
\eeq
This relation, summed over all $X \nu$ final states with $\lambda = \lambda_X - \lambda_\nu = +1/2$, implies that in \Eq{4}, $\Gamma_{\lambda = +1/2} = \Gamma_{\lambda = -1/2}$. This, in turn, implies that $\alpha = 0$. That is, the angular distribution is isotropic. It is to be emphasized that this isotropy is a consequence of CPT and rotational invariance alone. It does not depend on any further details of the interactions driving the decay.

For this isotropy in the Majorana case to be a useful probe of whether neutrinos are of Majorana or Dirac character, the decay angular distribution must be non-isotropic in the Dirac case. In contrast to the Majorana case, in the Dirac case the decay angular distribution does depend on the interaction. We assume that when $X = \gamma$, the decay is driven by effective neutrino transition magnetic and electric dipole moments $\mu$ and $d$. When $X = \pi^0$ or $\rho^0$, we take the decay to be dominated by a virtual $Z^0$ that emerges via a SM coupling from the neutrino line and becomes the $X$ particle. Finally, when $X = Z^0$, the $Z^0$ is simply emitted via a SM gauge coupling from the neutrino line, and when $X = H^0$, the $H^0$ is emitted via a Yukawa coupling from the neutrino line.

As desired, these processes do lead to non-isotropic angular distributions if neutrinos are Dirac particles. This is nicely illustrated by the decay $N \ra \nu + \pi^0$. Driven by an intermediate $Z^0$ exchange, this decay has, in the Dirac case, an amplitude proportional to 
\beq
\bar{u}_\nu /\hspace{-6pt}p_\pi \frac{(1-\gamma_5)}{2} u_N = m_N \left[ \frac{(1-\gamma_5)}{2} u_\nu\right] ^\dag \gamma^0 u_N \;\; .
\label{eq6}
\eeq
Here, $u_\nu$ and $u_N$ are Dirac wave functions for the neutrinos, $p_\pi$ is the pion momentum, and $m_N$ is the $N$ mass. So long as $m_N$ is not extremely close to the pion mass, the $\nu$ will be ultra-relativistic in the $N$ rest frame. Consequently, the left-handed chiral projection operator $(1-\gamma_5)/2$ acting on $u_\nu$ will allow only a $\nu$ of left-handed helicity to be emitted. That is, only decays with $\lambda \equiv \lambda_X - \lambda_\nu = \lambda_{\pi^0} - \lambda_\nu = + 1/2$ will be allowed. Hence, from \Eq{4}, in the Dirac case,
\beq
\frac{d \Gamma(N \ra \nu + \pi^0)}{d(\cos \theta)} \propto (1 + \cos \theta) \;\; .
\label{eq7}
\eeq
Given that the asymmetry parameter $\alpha$ in the decays $N \ra \nu + X$ must satisfy $ -1 \leq \alpha \leq +1$ (see \Eq{4}), the angular distribution of \Eq{7} is as far from isotropy as it is possible to get.

We find by explicit calculation that in the Dirac case, the asymmetry parameter $\alpha$ in the angular distribution
\beq
\frac{d \Gamma(N \ra \nu + X)}{d(\cos \theta)} =  \frac{\Gamma_0}{2} (1 + \alpha \cos \theta) 
\label{eq8}
\eeq
is as given in Table \ref{table1} for $X = \gamma, \; \pi^0, \; \rho^0, \; Z^0$, and $H^0$.
\begin{table}[ht]
\caption{The asymmetry parameter $\alpha$ in the angular distribution of the particle $X$ from the decay $N \ra \nu +X$ when $N$ and $\nu$ are Dirac particles. The quantities $m_N,\, m_\rho$ and $m_Z$ are the masses of the $N,\, \rho$, and $Z$, respectively. \vspace{6pt}}
\label{table1}
\Large
\begin{center}
\scalebox{1.2}  {
\begin{tabular}{|c||c|c|c|c|c|}
 \hline
X		& $\gamma$		& $\pi^0$	& $\rho^0$		& $Z^0$		& $H^0$	\\
\hline  
$\alpha$	&  $\frac{2\Im m (\mu d^*)}{|\mu|^2+|d|^2}$		& $1$	
		&  {\rule[-4mm]{0mm}{9mm} $\frac{m^2_N-2m_\rho^2}{m^2_N+2m_\rho^2} $ }
		&  $\frac{m^2_N-2m_Z^2}{m^2_N+2m_Z^2}$	& $1$	\\
\hline
\end{tabular}}  
\end{center}  
\end{table}
From Table \ref{table1} we see that, except in unlikely special circumstances such as $m_N^2 = 2m_\rho^2,\; \alpha$ is not zero. That is, the angular distribution is not isotropic, in contrast to its isotropy in the Majorana case. Moreover, once $m_N$ has been measured, the value of $\alpha$ in the Dirac case will be known for four of the five possible decays we have considered. Thus, the decay angular distributions of a heavy neutrino could be a quite fruitful probe of the Majorana vs.\ Dirac question.

Does a heavy neutrino actually exist? Such a neutrino is being sought at CERN \cite{ref2}. The potential for the Fermilab Short Baseline Neutrino program to discover such a neutrino through its decays has been considered by Ballett, Pascoli, and Ross-Lonergan \cite{ref3}. Some of the physics of such a neutrino has been discussed by Hernandez \etal  \, \cite{ref4} and by Caputo \etal \, \cite{ref5}.

In summary, we conclude that if a heavy neutrino is discovered, the angular distributions in its decays could tell us whether neutrinos are Dirac or Majorana particles.

\section*{Acknowledgments}
We thank Baha Balantekin and Andr\'{e} de Gouv\^{e}a for a very enjoyable collaboration. We are grateful to the organizers of the 53$^\mathrm{rd}$ Rencontres de Moriond session devoted to Electroweak Interactions and Unified Theories for many things.
This document was prepared using the resources of the Fermi National Accelerator Laboratory (Fermilab), a U.S. Department of Energy, Office of Science, HEP User Facility. Fermilab is managed by Fermi Research Alliance, LLC (FRA), under Contract No. DE-AC02-07CH11359.

\end{document}